\documentclass[%
 reprint,
 superscriptaddress,
 amsmath,amssymb,
 aps,
]{revtex4-2}

\usepackage{graphicx}  
\usepackage{dcolumn}   
\usepackage{bm}         
\usepackage{commath}
\usepackage[innercaption]{sidecap}
\usepackage[framemethod=TikZ]{mdframed}
\usepackage{pifont}
\usepackage{multirow}
\usepackage{amsthm}
\usepackage{mathrsfs}
\usepackage[title]{appendix}
\usepackage{xcolor}
\usepackage{xurl}
\usepackage{textcomp}
\usepackage{booktabs}
\usepackage{algorithm}
\usepackage{algorithmicx}
\usepackage{algpseudocode}
\usepackage{listings}
\usepackage{soul}
\usepackage{microtype}
\usepackage[colorlinks=true,citecolor=blue,linkcolor=blue,urlcolor=blue]{hyperref}

\makeatletter
\providecommand{\bstctlcite}[1]{%
  \@bsphack
  \@for\@citeb:=#1\do{%
    \edef\@citeb{\expandafter\@firstofone\@citeb}%
    \immediate\write\@auxout{\string\citation{\@citeb}}%
  }%
  \@esphack
}
\makeatother

\begin{document}

% \preprint{APS/123-QED}

%\title{Automated discovery in physics through machine-learning-guided systematic exploration}
\title{Large Language Model Based Agent for Automated Discovery \\in Computational Physics}

\author{Hang Lin}
\thanks{These two authors contributed equally to this work.}
 \affiliation{MOE Key Laboratory of Advanced Micro-Structured Materials, and School of Physical Science and Engineering, Tongji University, Shanghai, 200092, P. R. China}

\author{Chongwen Liu}
\thanks{These two authors contributed equally to this work.}
 \affiliation{MOE Key Laboratory of Advanced Micro-Structured Materials, and School of Physical Science and Engineering, Tongji University, Shanghai, 200092, P. R. China}

\author{Gang Yan}
\thanks{Contact author: gyan@tongji.edu.cn} 
\affiliation{MOE Key Laboratory of Advanced Micro-Structured Materials, and School of Physical Science and Engineering, Tongji University, Shanghai, 200092, P. R. China}
\affiliation{State Key Laboratory of Autonomous Intelligent Unmanned Systems, MOE Frontiers Science Center for Intelligent Autonomous Systems, and Shanghai Key Laboratory of Intelligent Autonomous Systems, Tongji University, Shanghai, 201210, P. R. China}

\date{\today}

\begin{abstract}
Scientific discovery in computational physics can often be framed as the optimization of quantitatively evaluable objectives subject to physical constraints. While researchers excel at formulating such problems, they frequently devote substantial effort to iterative refinement of methods and solution strategies. To accelerate this process, we introduce PhyNex, an autonomous agent that systematically explores the solution space of scorable scientific tasks by coupling large language model (LLM)-guided search with domain-specific computational tools that enforce physical consistency. PhyNex operates via progressive local search, accumulates reusable knowledge from both successful and failed attempts, and produces interpretable exploration trajectories that reveal which algorithmic components drive performance improvements. We validate PhyNex on three representative and scientifically important problems: predicting frequency-dependent dielectric spectra of semiconductors from crystal structure, designing probabilistic-circuit heuristics for Max-Cut on graphs, and optimizing charging protocols for Dicke quantum batteries in the chaotic coupling regime. Across the three tasks, PhyNex autonomously identifies solutions that match or exceed state-of-the-art approaches designed by human scientists, yielding search-averaged improvements of up to 3.8\% in spectral similarity, up to 15.0\% in normalized mean cut for Max-Cut, and 5.9\% in ergotropy at the $80\mathrm{k}$ training checkpoint in open exploration. These findings demonstrate that LLM-based agents with structured, feedback-driven exploration can substantially accelerate the path from problem specification to effective implementation, suggesting a practical division of labor in which scientists define objectives and constraints while automated systems navigate the methodological search space.
\end{abstract}

\maketitle

\section{\label{sec:intro}Introduction}

Many problems in scientific research can be abstracted as tasks characterized by explicit objectives, well-defined constraints, and quantitatively measurable evaluation criteria. Such tasks typically allow researchers to directly assess candidate solutions by specifying loss functions, error metrics, or experimental indicators. It is precisely these quantitatively scorable structures that enable scientific problems to be formulated systematically and provide a natural entry point for computational methods and automated exploration. For example, in materials science, one may seek to improve the energy density of batteries~\cite{hasan2025advancing}; in energy research, to achieve artificial photosynthesis with high solar-to-fuel conversion efficiencies~\cite{C3EE40831K,fehr2023integrated}; in fusion engineering, to stably increase the plasma gain factor~\cite{Shimada_2007}; in quantum computing, to surpass current limits of physical-gate fidelity~\cite{GoogleQuantumAI2023}; and in biomedicine, to develop biomarkers with high diagnostic accuracy~\cite{cao2023large}. What unites these disparate challenges is their scorability: each admits an objective metric against which candidate solutions can be evaluated. Nevertheless, identifying an effective solution remains a demanding endeavor. The process typically involves physical modeling, data processing, error diagnosis, and continual reflection and refinement -- a cycle that often requires months or even years of sustained effort. Early attempts to automate this cycle in the physical sciences have utilized modular neural-symbolic architectures to discover fundamental laws from raw data, effectively "snapping" complex observations into simple symbolic formulas~\cite{physicisttailin}.

{%
Large language models (LLMs) have progressed from exhibiting emergent abilities~\cite{wei2022emergent} to performing systematic reasoning and planning~\cite{wei2022chain,yao2023tree}. Researchers now routinely deploy them as assistants across diverse scientific disciplines~\cite{boiko2023autonomous,kang2024chatmof}. Beyond basic assistance, LLMs combined with automated evaluation have enabled evolutionary program-search methods. These methods yield algorithmic discoveries in mathematics and computer science~\cite{romera2024mathematical,novikov2025alphaevolve,fawzi2022alphatensor,mankowitz2023alphadev}. Concurrently, end-to-end research agents have begun to automate broader stages of scientific investigation. For example, ML-Master~\cite{liu2025ml,chan2024mle-bench} autonomously executes and debugs AI engineering tasks, and PhysMaster~\cite{miao2025physmaster} combines theoretical reasoning with numerical computation to close physics research loops. These advancements have expanded the scope of tasks that LLM-based agents can handle.}

{%
However, significant challenges remain~\cite{wang2023scientific,zhang2025_llm_scimethod}, and existing paradigms still leave substantial room for improvement.
Autonomous research agents are typically tailored to specific task classes, making adaptation to new domains costly and requiring substantial redesign. Evolutionary program-search methods provide greater flexibility, but their reliance on multiple simultaneous mutations obscures the causal contribution of individual modifications to performance gains. 
In computational physics, both aspects are critical. An effective system must not only generalize across diverse problems, but also provide interpretable attribution of performance improvements to specific algorithmic components. Such mechanistic insight is essential for guiding subsequent scientific discovery.}

{%
Here, we introduce PhyNex, an LLM-based agent designed for computational physics tasks with well-defined, scorable objectives. PhyNex explores the space of executable programs through localized, single-component modifications, enabling precise attribution of score variations to specific algorithmic choices. 
The resulting sequence of modifications, performance changes, and associated rationales forms an explicit exploration trajectory, revealing which design choices drive improvement and which lead to degradation, thereby providing interpretable physical insights.}

{%
We evaluate PhyNex on three representative problems: predicting frequency-dependent dielectric spectra of semiconductors from crystal structures~\cite{grunert2024deep}, designing probabilistic-circuit heuristics for Max-Cut on graphs~\cite{weitz2025subuniversal}, and optimizing charging protocols for Dicke quantum batteries in the chaotic coupling regime~\cite{erdman2024reinforcement}. 
In all cases, PhyNex identifies solutions that match or surpass human-designed baselines within 12 hours of computation, achieving improvements of 3.8\%, 15.0\%, and 5.9\%, respectively. 
Importantly, the generated trajectories exhibit physically interpretable patterns, including non-negative output constraints for optical absorption, degree-aware gate scheduling in scale-free networks, and quantile-based actor–critic strategies for quantum battery charging.}

\begin{figure*}[t]
  \centering
  \includegraphics[width=\textwidth]{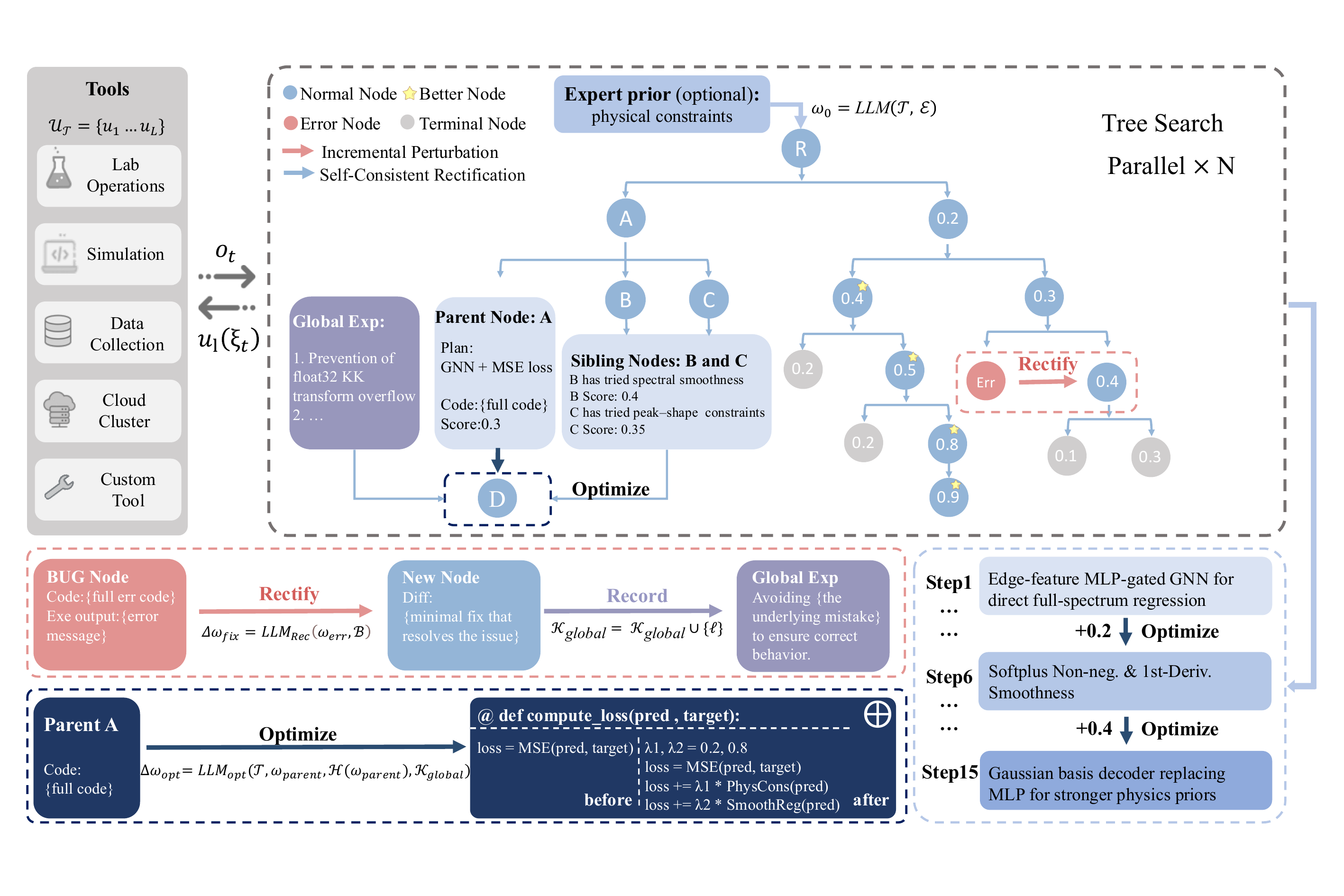}
  \caption{\label{fig:overview_framework}Architecture of the PhyNex framework that operates as a closed-loop agent. The process is initialized with a scientific task description (for example, predicting the full 0-20 eV optical spectrum of a material from its crystal structure) together with optional expert priors, while the left panel provides access to domain-specific computational tools. The central panel illustrates the parallel tree-search strategy used to explore the solution space, where each node represents a distinct executable code solution. The bottom panel depicts the iterative evolution process, which alternates between a rectification mechanism (Rectify, pink path) and a progressive optimization mechanism (Optimize, blue path).}
\end{figure*}

\section{\label{sec:method}Methodology}
{%
Developing an effective method for a computational physics 
task typically requires extensive effort in algorithm 
design, implementation, and debugging, even when the 
evaluation metric is well defined. PhyNex 
(Fig.~\ref{fig:overview_framework}) automates this 
process.}

\subsection{\label{sec:method-problem}Problem Formulation}

{%
Each task $\mathcal{T}$ consists of three components:}
\begin{equation}
{%
\mathcal{T} = (\mathcal{X},\; \mathcal{Y},\; \mathcal{U}),}
\label{eq:task}
\end{equation}
{%
where $\mathcal{X}$ is the input space, 
$\mathcal{Y}$ is the output space, and 
$\mathcal{U} = \{u_1, u_2, \dots, u_L\}$ is a set of 
domain-specific tools provided by the scientist.}

{%
A solution $\omega$ is an executable program that reads 
from $\mathcal{X}$, uses tools in $\mathcal{U}$, and 
produces output in $\mathcal{Y}$. We denote the feasible 
solution space by}
\begin{equation}
{%
\Omega_{\mathcal{T}} = \left\{ \omega \;\middle|\; 
\omega : \mathcal{X} \to \mathcal{Y} \right\}.}
\label{eq:solution_space}
\end{equation}

{%
Solutions interact with the task environment exclusively 
through function calls to $\mathcal{U}$. As illustrated 
in Fig.~\ref{fig:overview_framework}, these tools may 
include physical simulators, data loaders, and 
domain-specific evaluation routines. The scientist designs 
$\mathcal{U}$ to encode the constraints of the task, so 
that all candidate solutions automatically operate within 
the valid physical regime. In particular, $\mathcal{U}$ 
includes a scoring function $\mathcal{M}$ that assigns a 
numerical performance measure to each solution.}

{%
Our objective is to identify the highest-scoring solution 
in $\Omega_{\mathcal{T}}$:}
\begin{equation}
{%
\omega^{*} = \arg\max_{\omega \in \Omega_{\mathcal{T}}} 
\mathcal{M}(\omega).}
\label{eq:objective}
\end{equation}
{%
The space $\Omega_{\mathcal{T}}$ is discrete and the 
mapping $\omega \mapsto \mathcal{M}(\omega)$ is 
non-differentiable, so gradient-based optimization does 
not apply. PhyNex navigates this space by iteratively 
proposing, executing, and scoring candidate modifications.}

\subsection{\label{sec:progressive}Progressive Optimization 
with Knowledge Accumulation}

{%
PhyNex starts from an initial solution generated by the 
LLM from the task specification $\mathcal{T}$ and an 
optional exploration direction $\mathcal{E}$:}
\begin{equation}
{%
\omega_0 = \mathrm{LLM}(\mathcal{T},\, \mathcal{E}).}
\label{eq:initialization}
\end{equation}
{%
When provided, $\mathcal{E}$ directs the initial solution 
toward a specific methodological starting point (e.g., a 
particular algorithm family). When omitted, the LLM 
constructs $\omega_0$ from the task specification alone.}

{%
Starting from $\omega_0$, PhyNex refines the solution 
through a sequence of localized modifications. At each 
step, the LLM proposes a targeted change $\Delta\omega$ 
to one component of the current solution 
$\omega_{\mathrm{parent}}$:}
\begin{equation}
{%
\omega_{\mathrm{child}} 
  = \omega_{\mathrm{parent}} \oplus \Delta\omega,}
\label{eq:update}
\end{equation}
{%
where $\oplus$ denotes a local edit to the parent program. 
Restricting each step to a single component ensures that 
the resulting change in $\mathcal{M}$ can be attributed 
to a specific algorithmic choice.}

{%
Not all modifications execute successfully. When a 
candidate $\omega_{\mathrm{err}}$ produces a runtime 
error, the failing program and its diagnostic output 
$\mathcal{B}$ are passed to the LLM:}
\begin{equation}
{%
\Delta\omega_{\mathrm{rec}} 
  = \mathrm{LLM}_{\mathrm{rec}}(\omega_{\mathrm{err}},\, 
  \mathcal{B}).}
\label{eq:rectify}
\end{equation}
{%
The repaired candidate 
$\omega_{\mathrm{fix}} = \omega_{\mathrm{err}} \oplus 
\Delta\omega_{\mathrm{rec}}$ is then executed again. If 
the repair fails after three attempts, the branch is 
terminated.}

{%
Each successful repair yields a lesson $\ell$ that 
captures the failure mode and how to avoid it. A new 
lesson is added to $\mathcal{K}_{\mathrm{global}}$ only 
if the same failure pattern is not already represented. 
At each subsequent refinement step, all entries in 
$\mathcal{K}_{\mathrm{global}}$ are included in the LLM 
prompt. Because duplicates are filtered at insertion 
time, the knowledge base grows sublinearly with the 
number of failures, keeping the context cost bounded 
in practice.}

{%
To propose each modification, the LLM receives four 
inputs: the task specification, the current solution, 
the history of prior modifications from the same parent, 
and the accumulated failure lessons:}
\begin{equation}
{%
\Delta\omega
=
\mathrm{LLM}_{\mathrm{opt}}\!\left(
\mathcal{T},\,
\omega_{\mathrm{parent}},\,
\mathcal{H}(\omega_{\mathrm{parent}}),\,
\mathcal{K}_{\mathrm{global}}
\right),}
\label{eq:refinement}
\end{equation}
{%
where $\mathcal{H}(\omega_{\mathrm{parent}})$ records the 
modifications already attempted from the current parent, 
preventing redundant exploration.}

\subsection{\label{sec:search}Depth-Guided Parallel 
Exploration}

{%
The refinement loop described above evolves a single 
solution along one trajectory. In practice, PhyNex 
launches $K$ such trajectories in parallel, each modeled 
as an independent search tree $T^{(k)}$ rooted at a 
different initial solution $\omega_0^{(k)}$ generated by 
Eq.~(\ref{eq:initialization}).}

{%
Within each tree, the search follows a simple rule: 
accept a modification when it improves the score 
($\mathcal{M}(\omega_{\mathrm{child}}) > 
\mathcal{M}(\omega_{\mathrm{parent}})$) and continue 
from the improved solution; otherwise, terminate the 
branch. Branches that encounter non-recoverable 
execution failures are also terminated. Each tree 
therefore traces a monotonically improving path through 
the solution space.}

{%
Running multiple trees from different starting points 
reduces the likelihood of converging to a single local 
optimum. Because each initial solution $\omega_0^{(k)}$ 
may represent a qualitatively different methodological 
approach, parallel trees explore different algorithmic 
strategies for the same task. The trees are coupled 
through $\mathcal{K}_{\mathrm{global}}$, which propagates 
lessons across all branches, so that a failure encountered 
in one trajectory can be avoided in others.}

{%
PhyNex logs every modification alongside its score change, 
producing an exploration trajectory that can be examined 
post-hoc to identify which algorithmic choices drive 
performance gains.}

\section{\label{sec:results}Results}

We assess PhyNex across three domains: optical property prediction in materials, max-cut optimization on graphs, and control protocol synthesis for quantum batteries.{Unless otherwise noted, experiments are repeated three times to ensure the robustness of the results.}

\subsection{\label{sec:task1}Task 1: Spectral Prediction of Semiconductor Materials}

Recent advancements in artificial intelligence, particularly the integration of graph networks and optimization algorithms, have significantly accelerated the high-throughput discovery and structural prediction of functional materials\cite{cheng2022crystal}. Building on this capability, predicting the optical response directly from crystal structures has become essential for designing next-generation photovoltaics, photocatalysis, and optoelectronic devices. While the frequency-dependent dielectric function $\varepsilon(\omega)$ governs fundamental processes like absorption and dispersion, obtaining accurate spectra typically requires expensive first-principles calculations or time-consuming experiments\cite{rohlfing2000electron,runge1984density,aspnes1983dielectric}. This motivates a {scorable} task: predicting the full-frequency dielectric response directly from structural data\cite{grunert2024deep}.

\begin{figure*}[t]
\centering
\includegraphics[width=0.88\textwidth]{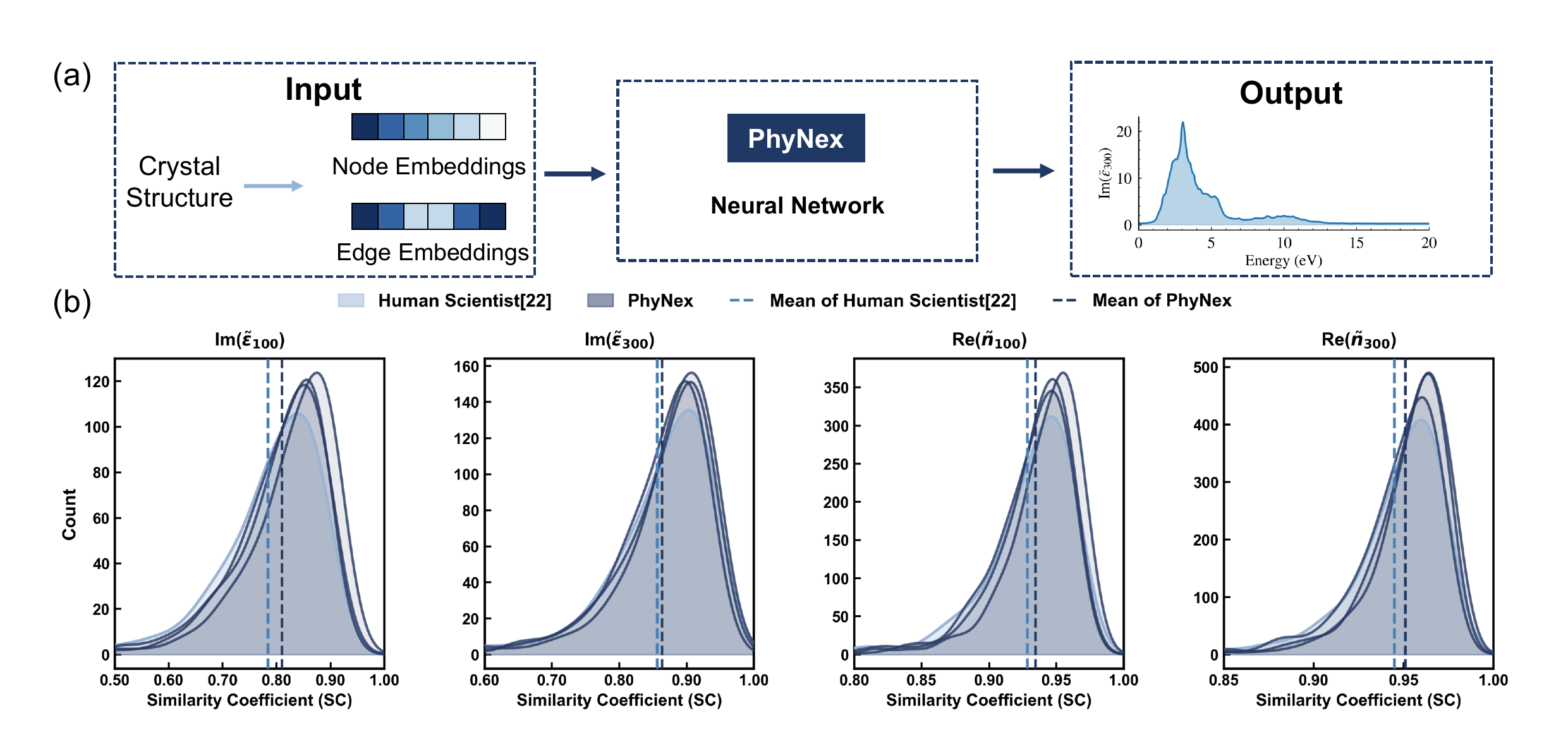}
\caption{\label{fig:task1_results}{Task~1: Spectral prediction of semiconductor materials. (a)~Schematic of the graph-to-spectrum prediction pipeline explored by PhyNex. (b)~Smoothed distributions of the test-set similarity coefficients across the four prediction targets. The Human Scientist distribution reports the baseline, while the PhyNex results are shown as three separate distributions corresponding to three independent searches. Dashed lines indicate the baseline mean and the pooled mean across the three PhyNex searches for each target.}}
\end{figure*}

{
We formalize this task as $\mathcal{T} = (\mathcal{X}, \mathcal{Y}, \mathcal{U})$, as illustrated schematically in Fig.~\ref{fig:task1_results}(a):

We identify the input space $\mathcal{X}$ as crystal structures represented as graphs. The specific node and edge embeddings follow Ref.~\cite{grunert2024deep} to ensure consistency with the original work. The output space $\mathcal{Y}$ comprises frequency dependent optical spectra discretized on an energy grid of 2001 equally spaced points spanning $0$ to $20~\mathrm{eV}$. Following the original work, we focus on the trace averaged dielectric function $\bar{\varepsilon} = \mathrm{Tr}(\varepsilon)/3$ and the corresponding averaged refractive index $\bar{n} = \sqrt{\bar{\varepsilon}}$. Four prediction targets are considered: the imaginary part of the dielectric function and the real part of the refractive index, each computed at two broadening parameters ($\eta = 100~\mathrm{meV}$ and $300~\mathrm{meV}$) to investigate the effect of spectral resolution on model performance. As for the tool set $\mathcal{U}$, the environment provides a minimum interface $\mathcal{U} = \{u_{\text{data}}, \mathcal{M}\}$. The data handling tool $u_{\text{data}}$ (\texttt{prepare\_dataloaders}) provides access to the official train, validation, and test splits with consistent graph representations. The scoring function $\mathcal{M}$ (\texttt{spectral\_similarity}) computes the similarity coefficient (SC) between the predicted spectrum $\tilde{Y}(\omega)$ and the reference spectrum $Y(\omega)$:
\begin{equation}
\mathcal{M}(\omega) = \mathrm{SC}\!\left[\tilde{Y}(\omega); Y(\omega)\right] = 1 - \frac{\int \left|Y(\omega) - \tilde{Y}(\omega)\right| d\omega}{\int \left|Y(\omega)\right| d\omega}.
\label{eq:task1_sc}
\end{equation}
This setup ensures all candidate solutions operate on identical data partitions and receive standardized performance feedback, isolating architectural exploration from data preparation.
}

\begin{table}[t]
\centering
\resizebox{\columnwidth}{!}{%
    \renewcommand{\arraystretch}{1.15}
    \begin{tabular}{lcccc}
    \hline
    Method & $\mathrm{Im}(\bar{\varepsilon}_{100})$ & $\mathrm{Im}(\bar{\varepsilon}_{300})$ & $\mathrm{Re}(\bar{n}_{100})$ & $\mathrm{Re}(\bar{n}_{300})$ \\
    \hline
    Human Scientist \cite{grunert2024deep} & 0.78 & 0.86 & 0.93 & 0.94 \\
    \textbf{PhyNex} & {$\mathbf{0.810 \pm 0.011}$} & {$\mathbf{0.863 \pm 0.006}$} & {$\mathbf{0.934 \pm 0.004}$} & {$\mathbf{0.951 \pm 0.003}$} \\
    \hline
    \end{tabular}%
}
\caption{\label{tab:task1_sc}Test-set similarity coefficient (Eq.~\ref{eq:task1_sc}) for Task~1 with $\mathrm{Im}\,\bar{\varepsilon}$ and $\mathrm{Re}\,\bar{n}$ evaluated under two broadening parameters ($\eta=100~\mathrm{meV}$ and $300~\mathrm{meV}$). Higher is better. {For PhyNex, we report the mean and standard deviation over three independent agent searches initialized from the same task specification and evaluated under an identical protocol.}}
\end{table}

{We now turn to the solutions discovered by PhyNex for this task. Starting from the task specification alone ($\mathcal{E} = \varnothing$), we launched PhyNex three times under the same evaluation interface. Across these independent searches, PhyNex discovers graph neural network solutions for spectral prediction. A representative architecture encodes crystal structures via three layers of edge-conditioned graph convolutions with batch normalization, aggregates node features through global mean pooling, and decodes the representation into a 2001-dimensional spectrum. PhyNex also autonomously introduced physically motivated output designs, including a Softplus activation for $\mathrm{Im}(\varepsilon)$ targets to enforce non-negative optical absorption and a baseline offset for refractive-index predictions to prevent unphysically low values.}

{Figure~\ref{fig:task1_results}(b) presents smoothed distributions of test-set similarity coefficients across all four prediction targets for the human-designed baseline together with three separate PhyNex distributions corresponding to three independent searches. The resulting PhyNex distributions exhibit a systematic rightward shift relative to the human baseline, and the low-accuracy tails are substantially suppressed, indicating that PhyNex can stably improve the mean performance with good robustness. As summarized in Table~\ref{tab:task1_sc}, the search-averaged SC reaches $0.810 \pm 0.011$ for $\mathrm{Im}(\bar{\varepsilon}_{100})$, $0.863 \pm 0.006$ for $\mathrm{Im}(\bar{\varepsilon}_{300})$, $0.934 \pm 0.004$ for $\mathrm{Re}(\bar{n}_{100})$, and $0.951 \pm 0.003$ for $\mathrm{Re}(\bar{n}_{300})$, consistently exceeding the human-designed baseline. Relative to the baseline values of $0.78$, $0.86$, $0.93$, and $0.94$, respectively, the largest average gain occurs on the challenging high-resolution imaginary dielectric function $\mathrm{Im}(\bar{\varepsilon}_{100})$, where the search-averaged SC increases to $0.81$. These results indicate that the physical constraints uncovered during autonomous exploration serve as effective inductive biases for spectral prediction.}

\subsection{\label{sec:task2}Task 2: Probabilistic-Circuit Max-Cut Optimization}

Max-Cut is a canonical NP-hard problem in combinatorial optimization: given a graph, the goal is to partition its vertices into two disjoint sets so as to maximize the number of edges crossing the partition. This problem is closely 
connected to finding the ground state of the Ising model in statistical physics. 
Classical approximation algorithms~\cite{goemans1995improved}
provide theoretical guarantees but face scalability challenges, while quantum 
approaches like QAOA~\cite{farhi2014quantum} remain limited by hardware noise. 
\emph{Classical probabilistic-circuit variational algorithms} offer an alternative 
by constructing variational ans\"atze using parameterized classical probability 
distributions~\cite{weitz2025subuniversal}. Task~2 aims to design such an algorithm 
with improved approximation performance.

\begin{figure*}[t]
\centering
\includegraphics[width=\textwidth]{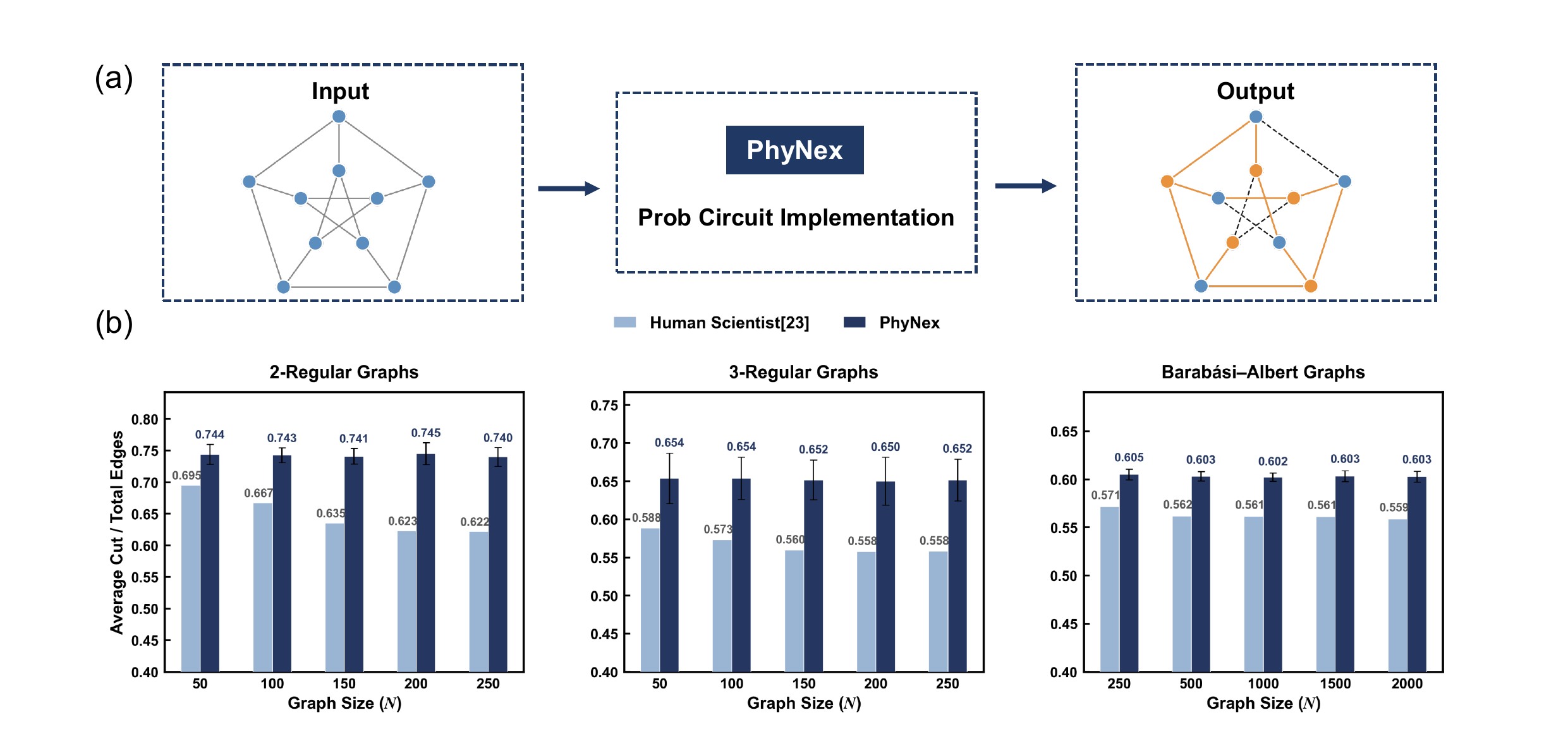}
\caption{{Task~2: Probabilistic-circuit Max-Cut optimization. (a)~Task schematic. (b)~Performance comparison across different graph families: 2-regular (left), 3-regular (middle), and Barab\'{a}si--Albert (right) graphs, using the normalized mean cut $\langle C\rangle / |E|$. The Human Scientist bars report the baseline means over 10 random instances per graph size. The PhyNex bars report the mean performance across three independent searches, with error bars indicating the standard deviation of these three means.}\label{fig:task2_results}}
\end{figure*}

{
We formalize this task as 
$\mathcal{T} = (\mathcal{X}, \mathcal{Y}, \mathcal{U})$, 
as illustrated in Fig.~\ref{fig:task2_results}(a). The 
input space $\mathcal{X}$ consists of graphs $G=(V,E)$ 
drawn from two families: regular graphs ($k$-regular with 
$k=2$ and $k=3$) and Barab\'asi--Albert (BA) scale-free 
networks with attachment parameter $m=2$. The output space 
$\mathcal{Y}$ is a binary string $z\in\{0,1\}^{n}$ 
indicating a bipartition of the $n=|V|$ vertices, with 
cut value
\begin{equation}
C(z) = \sum_{\langle ij\rangle \in E} 
\frac{1}{2}\Big[1 - (-1)^{z_i}(-1)^{z_j}\Big].
\label{eq:task2_cut}
\end{equation}

The tool set 
$\mathcal{U} = \{u_{\text{RE}}, u_{\text{BA}}\}$ provides 
graph-type-specific evaluation interfaces 
(\texttt{evaluate\_task\_RE} and 
\texttt{evaluate\_task\_BA}). Both tools enforce the 
structural constraints of the probabilistic circuit: 
execution proceeds via sequential two-bit probabilistic 
gates represented by valid stochastic matrices, and 
parameter optimization is handled internally using SPSA 
within a fixed budget of 100 iterations. The scoring 
function $\mathcal{M}$ is the normalized mean cut:
\begin{equation}
\mathcal{M} = \frac{\langle C\rangle_{\mathrm{est}}}{|E|},
\label{eq:task2_metric}
\end{equation}
where $\langle C\rangle_{\mathrm{est}}$ is the sample mean 
over repeated circuit runs. For regular graphs, the search phase optimizes 
$\mathcal{M}$ at $|V|=250$ on both 2-regular and 3-regular 
graphs; we then report performance on the standard 
benchmark sizes $N\in\{50,100,150,200,250\}$. For BA 
graphs, the search phase optimizes $\mathcal{M}$ at 
$|V|=250$; we then report performance on 
$N\in\{250,500,1000,1500,2000\}$, using $N=250$ as the 
search size and the larger graphs to assess generalization 
up to $|V|=2000$ vertices.
}

% At the tool level, PhyNex interacts with the task through interface 
% $\mathcal{U}_{\mathcal{T}}$ with two evaluation functions: 
% \texttt{evaluate\_task\_BA} for BA graphs and \texttt{evaluate\_task\_RE} 
% for regular graphs. Each function accepts a callback that constructs a 
% probabilistic circuit (a sequence of two-bit stochastic gates) from a 
% given graph and parameter vector. The evaluator handles SPSA optimization internally and returns the respective metrics~\cite{weitz2025subuniversal}.

{
We now describe the PhyNex solutions discovered for this task, presenting each graph family together with its corresponding results.
}

\textit{For Regular graphs,} {We launched PhyNex three independent times under the same task specification and evaluation protocol. One discovered regular-graph solution combines} layer-dependent adaptive sharpening with deterministic symmetry breaking. Each layer applies a learned exponent to sharpen column probabilities, producing an annealing-like transition from exploration to exploitation. Small biases indexed by edge ordering lift the degeneracy inherent in symmetric graphs. This compact solution uses only 4 parameters at depth $p=1$, independent of graph size. In contrast to R-PAOA~\cite{weitz2025subuniversal}, which enforces hard constraints forbidding transitions to non-cut states, PhyNex adopts soft biasing that permits such transitions with small probability, enabling escape from local optima in a manner analogous to finite-temperature annealing.

{
We evaluate on graph sizes $N\in\{50,100,150,200,250\}$, generating 10 random instances per size for both 2-regular and 3-regular families. Figure~\ref{fig:task2_results}(b) compares the human-designed baseline with the search-averaged PhyNex performance under the same evaluation protocol, with error bars indicating the standard deviation across three independent searches. Across both regular-graph families, the search-averaged PhyNex performance outperforms R-PAOA in normalized mean cut. Table~\ref{tab:task2_all} reports metrics averaged over 50 instances per family. In the search-averaged results, the family-averaged normalized mean cut increases from $0.649$ to $0.743$ on 2-regular graphs and from $0.567$ to $0.652$ on 3-regular graphs, while the mean per-instance standard deviation decreases from $4.8$ to $4.4$ and from $6.6$ to $6.2$, respectively. These gains are achieved by compact discovered regular-graph solutions using only 4 parameters, compared to hundreds required by R-PAOA (scaling as $|E|$, e.g., approximately 375 at $N=250$). The fixed parameter count also facilitates stable optimization and enables direct generalization across graph sizes without retuning.
}

\begin{table}[t]
\centering
\scriptsize
\setlength{\tabcolsep}{3pt}
\renewcommand{\arraystretch}{1.08}
\caption{{\label{tab:task2_all}Task~2 results for regular and Barab\'{a}si--Albert graphs. For PhyNex, we report the mean and standard deviation over three independent searches initialized from the same task specification and evaluated under an identical protocol. For regular graphs, the statistics are averaged over 50 instances per family; for BA graphs, they are averaged over the 50 test instances spanning the five evaluated sizes $N\in\{250,500,1000,1500,2000\}$.}}
\resizebox{\columnwidth}{!}{%
\begin{tabular}{l l c c c c}
\hline
Graph & Method & $\widehat{C}_{\max}$ & $\langle C\rangle$ & $\sigma$ & $\langle C\rangle / |E|$ \\
\hline
\multirow{2}{*}{2-Reg}
 & Human Scientist\cite{weitz2025subuniversal} & 107.8 & 95.4 & 4.8 & 0.649 \\
 & PhyNex & \textbf{122.0 $\pm$ 1.7} & \textbf{111.3 $\pm$ 2.7} & \textbf{4.4 $\pm$ 0.3} & \textbf{0.743 $\pm$ 0.018} \\
\hline
\multirow{2}{*}{3-Reg}
 & Human Scientist\cite{weitz2025subuniversal} & 144.0 & 126.5 & 6.6 & 0.567 \\
 & PhyNex & \textbf{162.0 $\pm$ 6.9} & \textbf{146.6 $\pm$ 7.9} & \textbf{6.2 $\pm$ 0.2} & \textbf{0.652 $\pm$ 0.036} \\
\hline
\multirow{2}{*}{BA}
 & Human Scientist\cite{weitz2025subuniversal} & 1454.7 & 1398.6 & 22.1 & 0.561 \\
 & PhyNex & \textbf{1563.6 $\pm$ 16.3} & \textbf{1505.0 $\pm$ 15.9} & \textbf{23.1 $\pm$ 0.4} & \textbf{0.603 $\pm$ 0.006} \\
\hline
\end{tabular}
}
\end{table}

\textit{For Barab\'{a}si--Albert graphs,} {We likewise launched PhyNex three independent times for BA graphs under the same task specification and evaluation protocol. One discovered BA solution} {exploits the heterogeneous degree distribution of BA graphs through three design choices. First, it prioritizes hub nodes: edges incident to vertices whose degree exceeds the 85th percentile are scheduled before the remaining edges, so that high-degree nodes (which influence many neighbors) are updated first. Second, it introduces temporal correlations between gates: each gate matrix is combined as a convex mixture with the most recently applied matrices at its two endpoints, creating a momentum-like effect that helps coordinate bit flips across layers. Third, it groups edges into three classes (hub--hub, hub--peripheral, peripheral--peripheral) and assigns a shared parameter set to each class, yielding 18 parameters at depth $p=1$ regardless of graph size. By contrast, R-PAOA assigns independent parameters to each edge (scaling as $|E|$), so PhyNex's fixed parameter count enables direct transfer to larger graphs without retuning.}

{
As BA graphs require larger sizes for their characteristic hub structure to emerge, we train PhyNex on graphs with $N=250$ and evaluate its performance on $N\in\{250,500,1000,1500,2000\}$ with 10 random instances per size, using the larger graphs to assess generalization beyond the search size. Figure~\ref{fig:task2_results}(b) reports the normalized mean cut $\langle C\rangle/|E|$ for the Barab\'{a}si--Albert family (right panel), while the PhyNex bars summarize the mean performance across three independent searches at each tested size, with error bars indicating the standard deviation of these three means; Table~\ref{tab:task2_all} further summarizes the BA results averaged over all 50 test instances. All three independently discovered PhyNex solutions outperform R-PAOA at every tested size. Averaged over all 50 BA test instances, the normalized mean cut increases from $0.561$ to $0.603$, and the aggregated bars indicate that this advantage remains stable as $N$ increases. These results confirm that PhyNex's hub-centric scheduling and temporally correlated updates effectively exploit the heterogeneous degree structure of scale-free networks, with the learned strategy generalizing beyond the training regime.
}

\subsection{\label{sec:task3}Task 3: Charging-Protocol Optimization for the Dicke Quantum Battery}
Conventional batteries exhibit charging times that scale linearly with capacity, fundamentally limiting power performance. Quantum batteries offer a promising alternative by exploiting collective quantum effects: theoretical studies have shown that charging time can decrease as $1/\sqrt{N}$ with battery size $N$, enabling a collective speedup~\cite{ferraro2018high,Campaioli2017PRL}. The Dicke quantum battery, comprising $N$ two-level systems (TLSs) coupled to a single-mode cavity, represents one of the most experimentally feasible designs. However, in the strong-coupling regime, quantum chaotic dynamics generates strong correlations and drives local subsystems toward thermal-like states, severely suppressing the extractable energy (ergotropy) to below 30\% under standard protocols~\cite{erdman2024reinforcement}. This motivates a {scorable} task: discovering time-dependent control protocols that counter quantum chaos and maximize energy extraction.

\begin{figure*}[t]
\centering
\includegraphics[width=\textwidth]{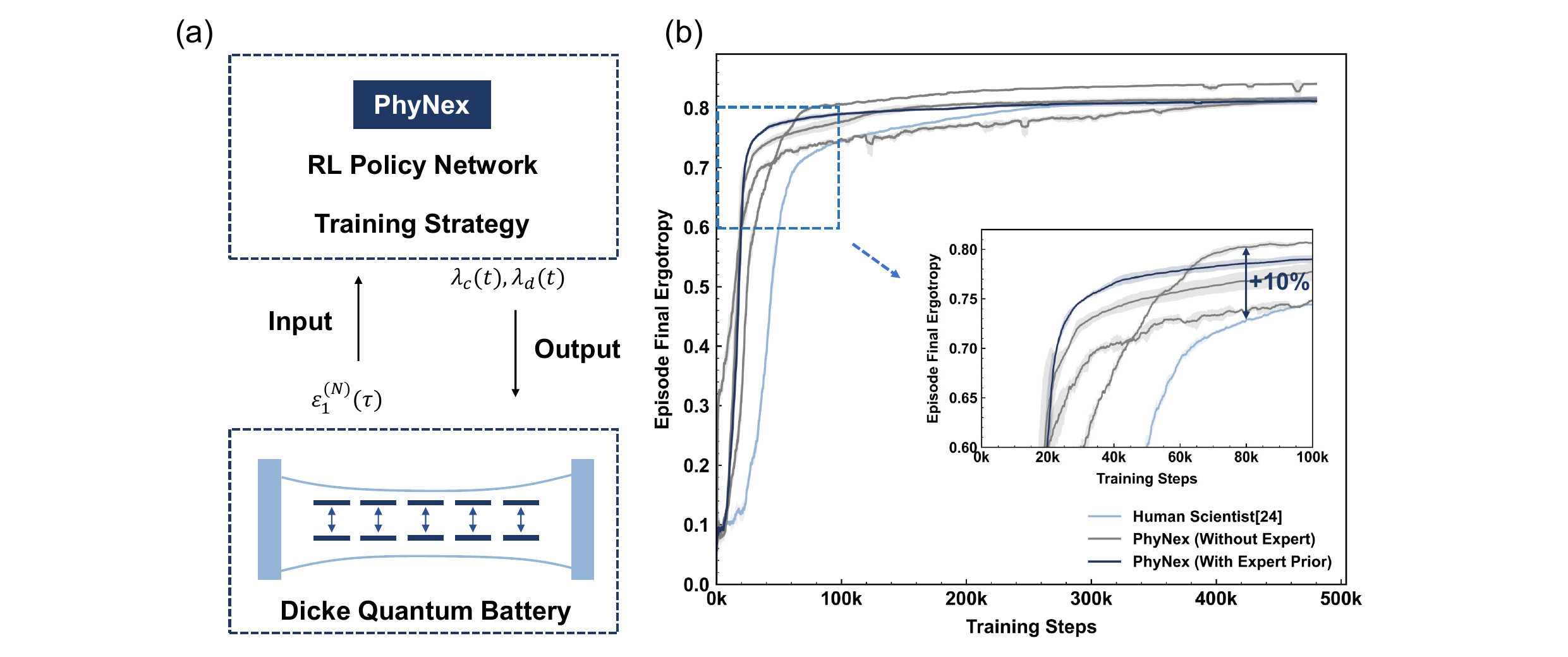}
\caption{\label{fig:task3_results}{Task~3: Charging-protocol optimization for the Dicke quantum battery. (a) Schematic of the optimization framework. (b) Convergence curves of episode final ergotropy for the Human Scientist baseline, PhyNex without expert prior, and PhyNex with expert prior; the inset zooms in on the first 100k training steps and highlights the 80k-step evaluation checkpoint, at which the single best PhyNex trajectory among the with- and without-expert runs achieves about 10\% higher ergotropy than the Human Scientist baseline.}}
\end{figure*}

{
We adopt the coupling-modulation subtask from 
Ref.~\cite{erdman2024reinforcement} and formalize it as 
$\mathcal{T} = (\mathcal{X}, \mathcal{Y}, \mathcal{U})$, 
as illustrated in Fig.~\ref{fig:task3_results}(a). The 
input space $\mathcal{X}$ is a fixed experimental 
configuration: $N=16$ two-level systems, effective coupling 
strength $\widetilde{g}=0.3\omega_0$ (within the chaotic 
regime), charging time $\tau=5.6\,\omega_0^{-1}$, and an 
initial state combining the battery ground state 
$|\mathrm{G}\rangle=\bigotimes_{j=1}^{N}|0\rangle_j$ with 
the $N$-photon Fock state $|N\rangle$ in the cavity. The 
output space $\mathcal{Y}$ is a piecewise-constant control 
protocol $\lambda_c(t) \in [-0.3,\, 0.3]$ with time step 
$\Delta t = 0.2\,\omega_0^{-1}$, yielding 28 discrete 
control actions over the charging period.

The tool set $\mathcal{U} = \{u_{\text{env}}, \mathcal{M}\}$ follows the interface defined in Ref.~\cite{erdman2024reinforcement}. The simulator $u_{\text{env}}$ (\texttt{make\_coupling\_environment}) evolves the system under the time-dependent Dicke Hamiltonian and switches off the coupling after charging. The scoring function $\mathcal{M}$ (\texttt{evaluate\_coupling\_policy}) returns the per-unit ergotropy at training step $K=80\mathrm{k}$:
\begin{equation}
\mathcal{M} = \mathcal{E}_1^{(N)}(\tau)\big|_{k=K} = \left( \frac{E^{(N)}(\tau)}{N} - r_1(\tau)\omega_0 \right)\Bigg|_{k=K},
\label{eq:task3_M}
\end{equation}
where $E^{(N)}(\tau) = \langle\psi(\tau)|\widehat{\mathcal{H}}_{\mathrm{B}}|\psi(\tau)\rangle$ is the mean battery energy and $r_1(\tau)$ is the minimum eigenvalue of the single-TLS reduced density matrix. Evaluating at this fixed $80\mathrm{k}$-step checkpoint measures converged performance while reducing the computational cost during iterative exploration. For the final performance comparison against the Human Scientist baseline, the discovered policy is evaluated at $480\mathrm{k}$ steps, following the protocol of the original work.

To isolate the effect of algorithmic priors, we evaluate 
PhyNex under two initialization modes: \textit{open 
exploration} ($\mathcal{E} = \varnothing$), starting from 
the task description alone; and \textit{guided exploration} 
($\mathcal{E} = \mathrm{SAC}$), providing a minimal prior 
that suggests the Soft Actor-Critic algorithm as a starting 
point without implementation details.
}

{The two initialization modes serve as an ablation study: by comparing guided exploration ($\mathcal{E}=\mathrm{SAC}$) with open exploration ($\mathcal{E}=\varnothing$), we isolate the effect of providing an expert algorithmic prior on both sample efficiency and solution diversity. We first describe the solutions discovered under each mode, then compare their performance quantitatively.}

{In guided exploration ($\mathcal{E}=\mathrm{SAC}$), PhyNex starts from the same SAC prior that underlies the human-designed baseline, but adapts it to the chaotic regime through two principal modifications. First, prioritized experience replay is replaced with uniform sampling, reducing overfitting to transitions with large TD errors, which in this setting often reflect the inherent noise of quantum dynamics rather than poor policy choices. Second, an auxiliary smoothness penalty is introduced into the actor loss, penalizing the squared difference between consecutive actions. This regularization discourages rapid fluctuations in the control signal $\lambda_c(t)$, promoting smooth control sequences. This smoothness is physically motivated: in chaotic quantum systems, abrupt changes in $\lambda_c$ can excite unwanted transitions and destabilize charging.}

{In open exploration ($\mathcal{E}=\varnothing$), we launched PhyNex three independent times under the same task interface. These independent searches produced multiple distinct reinforcement-learning strategies rather than converging to a single solution. One high-performing solution belongs to the deterministic actor--critic family and is augmented with a distributional quantile critic (QR-DDPG). The critic learns the full return distribution rather than only its mean, and the actor optimizes against a lower quantile of this distribution, favoring charging protocols that perform reliably even when returns are uncertain. The discovered algorithm further combines this quantile-based objective with a staged exploration-noise schedule, enabling broad exploration early in training and stable low-noise refinement at later stages.}

{We next compare the effects of providing an expert prior. Figure~\ref{fig:task3_results}(b) presents the training dynamics of the human-designed baseline, the guided PhyNex variant, and the three no-prior PhyNex searches. When expert prior knowledge is provided, PhyNex discovers a SAC-based variant that attains substantially higher ergotropy than the human-designed method throughout the early-training regime and at the $80\mathrm{k}$ checkpoint, while reaching a slightly lower value at $480\mathrm{k}$. Without prior knowledge, PhyNex nevertheless discovers multiple competitive solutions. As summarized in Table~\ref{tab:task3_results}, the search-averaged no-prior performance also exceeds the baseline at the $80\mathrm{k}$ checkpoint. The ``+10\%'' label in Fig.~\ref{fig:task3_results}(b) marks the approximate gain of the single best PhyNex trajectory at the $80\mathrm{k}$ checkpoint relative to the Human Scientist baseline.}

{Table~\ref{tab:task3_results} summarizes quantitative results. At the $80\mathrm{k}$ checkpoint, which defines PhyNex's search objective $\mathcal{M}$, the guided mode ($\mathcal{E}=\mathrm{SAC}$) achieves an ergotropy of $0.787$, a $7.78\%$ improvement over the baseline ($0.730$), while reaching $0.8124$ at $480\mathrm{k}$, slightly below the baseline value of $0.8183$. For open exploration, we report the mean and standard deviation over three independent no-prior searches, yielding $0.7733 \pm 0.0222$ at $80\mathrm{k}$ and $0.8221 \pm 0.0138$ at $480\mathrm{k}$, corresponding to a mean improvement of $5.90\%$ at the search checkpoint. These search-averaged no-prior results exceed the human-designed baseline at the $80\mathrm{k}$ checkpoint and remain slightly above it at $480\mathrm{k}$. Taken together, the results indicate that expert prior knowledge improves sample efficiency, while open exploration broadens the space of viable algorithmic solutions and yields robust long-horizon performance.}

\begin{table}[t]
\centering
\scriptsize
\setlength{\tabcolsep}{5pt}
\renewcommand{\arraystretch}{1.15}
\begin{tabular}{lccc}
\hline
Method & $\mathcal{M}_{80\mathrm{k}}$ & $\mathcal{M}_{480\mathrm{k}}$ & $\Delta_{80\mathrm{k}}$ \\
\hline
Human Scientist~\cite{erdman2024reinforcement} & 0.7302 & 0.8183 & +0.00\% \\
{PhyNex ($\mathcal{E}=\varnothing$)} & {$0.7733 \pm 0.0222$} & {$\mathbf{0.8221 \pm 0.0138}$} & {+5.90\%} \\
PhyNex ($\mathcal{E}=\mathrm{SAC}$) & \textbf{0.7870} & 0.8124 & \textbf{+7.78\%} \\
\hline
\end{tabular}
\caption{\label{tab:task3_results}Task~3 performance comparison. $\mathcal{M}_{k}$ denotes the per-unit ergotropy at training step $k$. $\Delta_{80\mathrm{k}}$ indicates the relative improvement over the baseline at the search checkpoint. {For open exploration, we report the mean and standard deviation over three independent no-prior PhyNex searches initialized from the same task specification and evaluated under an identical protocol. Guided exploration ($\mathcal{E}=\mathrm{SAC}$) was evaluated with a single run, so no standard deviation is reported.}}
\end{table}

\section{\label{sec:discussion}Discussion}

\subsection{{What Exploration Trajectories Reveal}}

{%
Across all three tasks, PhyNex identified solutions that
matched or exceeded expert-designed methods within
12 hours and at a cost on the order of \$5. The discovered
methods are physically interpretable: non-negative output
constraints for optical absorption in Task~1, degree-aware
gate scheduling for scale-free graphs in Task~2, and both
temporal smoothness regularization and conservative
quantile-based control for quantum charging in Task~3.}

{%
Beyond the final solutions, the exploration trajectories
enable attribution of performance changes to specific
algorithmic choices. Because each modification targets a
single component with a recorded score change, a researcher
can read a trajectory to answer concrete diagnostic
questions: which design choice produced the largest gain,
which degraded performance, and whether two modifications
interact. In Task~1, for example, enforcing non-negative
absorption via a Softplus activation yielded the largest
single gain in spectral similarity, immediately identifying
output positivity as the most important inductive bias for
this architecture. In Task~3, the guided branch shows that
replacing prioritized experience replay with uniform
sampling and adding a smoothness penalty each produced a
distinct measurable improvement, separating two
modifications that a researcher would otherwise have to
test independently by hand. The open-exploration branch
identified a QR-DDPG-style solution in which a
conservative quantile objective and a staged
exploration-noise schedule together yield stronger
long-horizon performance. Extracting such cause--effect
patterns across tasks could inform the design of related
computational physics algorithms. For a comprehensive empirical
demonstration, we refer readers to the Supplementary
Information (Sec.~V), which details a stepwise case study
of a 7-iteration evolutionary trajectory isolating the
dynamic operators in the regular-graph Max-Cut problem.}

\subsection{Scope and Limitations}
PhyNex addresses a specific class of problems: computational tasks with 
{scorable} objectives and moderate per-evaluation cost. This 
scope covers substantial territory in computational physics, from numerical 
algorithm design to protocol optimization, but excludes tasks where 
evaluation is prohibitively expensive (large-scale DFT, high-precision 
many-body calculations) or where no quantitative metric captures the 
scientific goal (proposing theoretical frameworks, interpreting anomalous 
observations).

Two design choices shape the framework's behavior. First, exploration 
proceeds through local modifications rather than global restructuring. 
This favors interpretability and safety, but may miss qualitatively 
different solution families requiring non-local jumps—a limitation when 
such alternatives exist, irrelevant when they do not. Second, exploration 
is inherently stochastic: independent runs may arrive at distinct solutions. 
We view this as a feature. The multiplicity of discovered methods reflects 
the problem landscape's structure and can reveal alternative mechanisms 
worth investigating. Complete trajectory logging ensures any specific run 
remains reproducible.

{%
Because the search objective in Task 3 is defined at the $80\mathrm{k}$ 
checkpoint, the reported gains are naturally most pronounced there; future 
work should assess robustness to alternative evaluation checkpoints. In 
addition, the guided-exploration result in Task 3 is based on a single run 
and should therefore be interpreted as preliminary rather than as a robust 
estimate of variance-sensitive performance.}

A practical note on task specification: effective use of
PhyNex does not require elaborate prompt engineering.
{%
None of the three tasks in this study required more than a
plain-language description of the objective, the
constraints, and the evaluation metric.} The essential
requirement is clarity of formulation, including what
constitutes a valid solution, what constraints must be
respected, and how quality is measured, rather than
sophistication of presentation.
{%
The main human effort lies in designing the tool interface
$\mathcal{U}$, which encodes the physical constraints and
evaluation protocol. This effort, however, coincides with
the problem-formulation stage that any computational study
already requires. PhyNex does not replace physical
insight; it automates the subsequent trial-and-error loop
of algorithm implementation and hyperparameter tuning,
allowing the researcher to concentrate on formulating
faithful constraints and objective functions.}

\subsection{Future Directions}
The present work validates the core concept: once a problem can be cast as a scorable computational task with explicit physical constraints, systematic exploration can be automated effectively. The natural next step is extending this paradigm to broader scenarios.

The natural extension is integration with physical experiments. All evaluations reported here are computational, but the framework's architecture, including structured exploration, automatic rectification, and knowledge accumulation, is not inherently limited to simulation. Connecting PhyNex to laboratory platforms would enable direct optimization of experimental protocols, from quantum control sequences to spectroscopic measurements, with feedback drawn from real physical systems rather than numerical models.

Two further extensions merit investigation. First, the knowledge accumulated during exploration, particularly failure modes and successful corrections, is currently scoped to individual tasks. Abstracting this experience for reuse across related problems could transform PhyNex from a task-specific tool into a progressively more capable research assistant. Second, realistic research involves multi-stage workflows rather than isolated tasks; extending the framework to coordinate sequential subproblems would bring it closer to actual scientific practice. 

%How far can this paradigm be pushed? Our results suggest the boundaries lie not in the automation itself, but in the clarity with which problems can be specified and evaluated.

\begin{acknowledgments}
This work was supported by the National Natural Science Foundation of China (Grant No.~T2225022 and No.~62088101), the Shanghai Municipal Science and Technology Major Project (Grant No.~2021SHZDZX0100), and the Artificial Intelligence Empowering Disciplines Initiative of Shanghai Municipal Education Commission.
%We also thank DP Technology for providing computational resources.
\end{acknowledgments}

\section*{data availability}
The source code implementing the agentic framework PhyNex, along with the scripts to reproduce the figures and analysis presented in this study, is available on GitHub at \url{https://github.com/lhhhappy/PhyNex}. 

\bibliographystyle{apsrev4-2-titleit} % PRR 推荐的样式
\bstctlcite{apsrev4-2Control}
\bibliography{refs_updated}           % 对应 refs.bib

\end{document}